\documentclass[12pt]{article}
\usepackage{latexsym}
\usepackage{amsmath,amsfonts}
\usepackage{times}

\catcode`\@=11

\newcount\hour
\newcount\minute
\newtoks\amorpm \hour=\time\divide\hour by 60\minute
=\time{\multiply\hour by 60 \global\advance\minute by-\hour}
\edef\standardtime{{\ifnum\hour<12 \global\amorpm={am}%
        \else\global\amorpm={pm}\advance\hour by-12 \fi
        \ifnum\hour=0 \hour=12 \fi
        \number\hour:\ifnum\minute<10
        0\fi\number\minute\the\amorpm}}
\edef\militarytime{\number\hour:\ifnum\minute<10 0\fi\number\minute}

\def\draftlabel#1{{\@bsphack\if@filesw {\let\thepage\relax
   \xdef\@gtempa{\write\@auxout{\string
      \newlabel{#1}{{\@currentlabel}{\thepage}}}}}\@gtempa
   \if@nobreak \ifvmode\nobreak\fi\fi\fi\@esphack}
        \gdef\@eqnlabel{#1}}
\def\@eqnlabel{}
\def\@vacuum{}
\def\marginnote#1{}
\def\draftmarginnote#1{\marginpar{\raggedright\scriptsize\tt#1}}
\def\draft{
        \pagestyle{plain}
        \overfullrule=2pt
        \oddsidemargin -.5truein
        \def\@oddhead{\sl \phantom{\today\quad\militarytime} \hfil
        \smash{\Large\sl DRAFT} \hfil \today\quad\militarytime}
        \let\@evenhead\@oddhead
        \let\label=\draftlabel
        \let\marginnote=\draftmarginnote
        \def\ps@empty{\let\@mkboth\@gobbletwo
        \def\@oddfoot{\hfil \smash{\Large\sl DRAFT} \hfil}
        \let\@evenfoot\@oddhead}
        \def\@eqnnum{(\theequation)\rlap{\kern\marginparsep\tt\@eqnlabel}%
        \global\let\@eqnlabel\@vacuum}  }

\newcommand{\rf}[1]{(\ref{#1})}
\renewcommand{\theequation}{\thesection.\arabic{equation}}
\renewcommand{\thefootnote}{\fnsymbol{footnote}}
\newcommand{\newsection}{   
\setcounter{equation}{0}\section}

\def\appendix#1{\addtocounter{section}{1}\setcounter{equation}{0}
\renewcommand{\thesection}{\Alph{section}}
\section*{Appendix \thesection\protect\indent \parbox[t]{11.15cm}{#1}}
\addcontentsline{toc}{section}{Appendix \thesection\ \ \ #1}}

\def\be{\begin{equation}}
\def\ee{\end{equation}}
\def\beq{\begin{eqnarray}}
\def\eeq{\end{eqnarray}}

\def\parline{\,\partial\kern -0.55em /\,\,}

\def\half{{\frac{1}{2}}}

\def\AA{{\cal A}}

\def\EE{{\cal E}}

\def\LL{{\cal L}}
\def\MM{{\cal M}}
\def\NN{{\cal N}}

\def\Cbf{{\bf C}}
\def\Dbf{{\bf D}}

\def\phik{|\phi\rangle}
\def\phibr{\langle\phi|}

\def\Phik{|\Phi\rangle}
\def\Phibr{\langle\Phi|}

\def\xik{|\xi\rangle}
\def\Xik{|\Xi\rangle}

\def\smzero{{\scriptscriptstyle (0)}}
\def\smone{{\scriptscriptstyle (1)}}
\def\smtwo{{\scriptscriptstyle (2)}}

\def\smx#1{{\scriptscriptstyle (#1)}}

\def\smzero{{\scriptscriptstyle (0)}}
\def\smone{{\scriptscriptstyle (1)}}
\def\smtwo{{\scriptscriptstyle (2)}}

\def\smx#1{{\scriptscriptstyle (#1)}}

\def\smpone{{\scriptscriptstyle [1]}}

\def\smponetwo{{\scriptscriptstyle [1,2]}}

\def\smptwothree{{\scriptscriptstyle [2,3]}}

\def\Cwt{\widetilde{C}}

\def\alpar{\alpha\partial}
\def\albpar{\bar\alpha\partial}

\def\eb{\bar{e}}
\def\mb{\bar{m}}

\def\alphabf{{\boldsymbol{\alpha}}}

\begin{document}


\begin{flushright}
FIAN-TD-2008-17 \qquad \ \ \ \ \ \\
arXiv: 0808.3945 [hep-th]
\end{flushright}

\vspace{1cm}

\begin{center}

{\Large \bf CFT adapted gauge invariant formulation of arbitrary spin

\medskip
fields in AdS and modified de Donder gauge}

\vspace{2.5cm}

R.R. Metsaev\footnote{ E-mail: metsaev@lpi.ru }

\vspace{1cm}

{\it Department of Theoretical Physics, P.N. Lebedev Physical
Institute, \\ Leninsky prospect 53,  Moscow 119991, Russia }

\vspace{3.5cm}

{\bf Abstract}

\end{center}

Using Poincar\'e parametrization of AdS space, we study totally symmetric
arbitrary spin massless fields in AdS space of dimension greater than or
equal to four. CFT adapted gauge invariant formulation for such fields is
developed. Gauge symmetries are realized similarly to the ones of
Stueckelberg formulation of massive fields. We demonstrate that the curvature
and radial coordinate contributions to the gauge transformation and
Lagrangian of the AdS fields can be expressed in terms of ladder operators.
Realization of the global AdS symmetries in the conformal algebra basis is
obtained. Modified de Donder gauge leading to simple gauge fixed Lagrangian
is found. The modified de Donder gauge leads to decoupled equations of motion
which can easily be solved in terms of the Bessel function. Interrelations
between our approach to the massless AdS fields and the Stueckelberg approach
to massive fields in flat space are discussed.

\newpage
\renewcommand{\thefootnote}{\arabic{footnote}}
\setcounter{footnote}{0}

\section{Introduction}

Further progress in understanding AdS/CFT correspondence
\cite{Maldacena:1997re} requires, among other things, better understanding of
field dynamics in $AdS$ space. Conjectured duality  of conformal SYM theory
and superstring theory in $AdS_5 \times S^5$ has lead to intensive and
in-depth study of various aspects of $AdS$ field dynamics. Although many
interesting approaches to $AdS$ fields are known in the literature (for
review see \cite{Bekaert:2005vh}-\cite{Fotopoulos:2008ka}), analysis of
concrete dynamical aspects of such fields is still a challenging procedure.
One of ways to simplify analysis of field and string dynamics in $AdS$ space
is based on use of the Poincar\'e parametrization of
$AdS$ space%
\footnote{ Studying $AdS_5\times S^5$ superstring action
\cite{Metsaev:1998it} in Poincar\'e parametrization may be found in
Ref.\cite{Metsaev:2000ds}. Recent interesting application of Poincar\'e
coordinates to studying $AdS_5\times S^5$ string $T-$duality  may be found in
\cite{Beisert:2008iq} (see also \cite{Berkovits:2008ic}).}.
Use of the Poincar\'e coordinates simplifies analysis of many aspect of $AdS$
field dynamics and therefore these coordinates have extensively been used for
studying the AdS/CFT correspondence. In this paper we develop a formulation
which  is based on considering of $AdS$ field dynamics in the Poincar\'e
coordinates. This is to say that using the Poincar\'e parametrization of
$AdS$ space we discuss massless totally symmetric arbitrary spin-$s$, $s\geq
1$, bosonic field propagating in $AdS_{d+1}$ space of dimension $d+1 \geq 4$.
Our results can be summarized as follows.

i) Using the Poincar\'e parametrization of $AdS$, we obtain gauge invariant
Lagrangian for free massless arbitrary spin $AdS$ field.  The Lagrangian is
{\it explicitly invariant with respect to boundary Poincar\'e symmetries},
i.e., manifest symmetries of our Lagrangian are adapted to manifest
symmetries of boundary CFT. We show that all the curvature and radial
coordinate contributions to our Lagrangian and gauge transformation are
entirely expressed in terms of ladder operators that depend on radial
coordinate and radial derivative. Besides this, our Lagrangian and gauge
transformation are similar to the ones of Stueckelberg formulation of massive
field in flat $d$-dimensional space. General structure of the Lagrangian we
obtained is valid for any theory that respects Poincar\'e symmetries. Various
theories are distinguished by appropriate ladder operators.

ii) We find modified de Donder gauge that leads to simple gauge fixed
Lagrangian. The surprise is that this gauge gives {\it
decoupled equations of motion}%
\footnote{ Our modified de Donder gauge seems to be unique first-derivative
gauge that leads to decoupled equations of motion. Light-cone gauge
\cite{Metsaev:1999ui} also leads to decoupled equations of motion, but the
light-cone gauge breaks boundary Lorentz symmetries.}.
Note that the standard de Donder gauge leads to coupled equations of motion
whose solutions for $s\geq 2$ are not known in closed form so far. In
contrast to this, our modified de Donder gauge leads to simple decoupled
equations which are easily solved in terms of the Bessel function.
Application of our approach to studying the AdS/CFT correspondence may be
found in Ref.\cite{Metsaev:2008fs}.

Motivation for our study of higher-spin $AdS$ fields in Poincar\'e
parametrization which is beyond the scope of this paper may be found at the
end of Section 5.

\newsection{ Lagrangian and gauge symmetries}

We begin with discussion of field content of our approach. In
Ref.\cite{Fronsdal:1978vb}, the massless spin-$s$ field propagating in
$AdS_{d+1}$ space is described by double-traceless $so(d,1)$
algebra totally symmetric tensor field $\Phi^{A_1\ldots A_s}$.%
\footnote{ $A,B,C=0,1,\ldots, d$ and $a,b,c=0,1,\ldots, d-1$ are the
respective flat vector indices of the $so(d,1)$ and $so(d-1,1)$ algebras. In
Poincar\'e parametrization of $AdS_{d+1}$ space, $ds^2=(dx^adx^a +
dzdz)/z^2$. We use the conventions: $\partial_a \equiv\partial/\partial x^a$,
$\partial_z\equiv\partial/\partial z$. Vectors of $so(d,1)$ algebra are
decomposed as $X^A=(X^a,X^z)$.}
This tensor field can be decomposed in scalar, vector, and totally symmetric
tensor fields of the $so(d-1,1)$ algebra:
\be \label{18052008-02}
\phi_{s'}^{a_1\ldots a_{s'}}\,, \hspace{2cm} s'=0,1,\ldots,s-1,s.
\ee
The fields $\phi_{s'}^{a_1\ldots a_{s'}}$ with $s'>3$ are double-traceless%
\footnote{ Note that $so(d-1,1)$ tensorial components of the Fronsdal field
$\Phi^{A_1\ldots A_s}$ are not double-traceless. Using appropriate
transformation (see \rf{05012008}) those tensorial components can be
transformed to our fields in \rf{18052008-02}.}
\be \label{18052008-03} \phi_{s'}^{aabba_5\ldots a_{s'}}=0\,, \hspace{2cm}
s'=4,5,\ldots,s-1,s. \ee
The fields in \rf{18052008-02} subject to constraints \rf{18052008-03}
constitute a field content of our approach. To simplify presentation we use a
set of the creation operators $\alpha^a$, $\alpha^z$, and the respective set
of annihilation operators, $\bar{\alpha}^a$, $\bar{\alpha}^z$. Then, fields
\rf{18052008-02}
can be collected into a ket-vector $|\phi\rangle$ defined by%
\footnote{ We use oscillator formulation
\cite{Lopatin:1987hz}-\cite{Labastida:1987kw} to handle the many indices
appearing for tensor fields (see also \cite{Bekaert:2006ix}). It can also be
reformulated as an algebra acting on the symmetric-spinor bundle on the
manifold $M$ \cite{Hallowell:2005np}.}
\beq
\label{18052008-04} &&  |\phi\rangle \equiv \sum_{s'=0}^s
(\alpha^z)^{s-s'}|\phi_{s'}\rangle \,,
\\
\label{18052008-05} && |\phi_{s'}\rangle \equiv
\frac{1}{s'!\sqrt{(s - s')!}}\,
\alpha^{a_1} \ldots \alpha^{a_{s'}} \, \phi_{ s'}^{a_1\ldots a_{s'}}
|0\rangle\,.
\eeq
From \rf{18052008-04},\rf{18052008-05}, we see that the ket-vector
$|\phi\rangle$ is degree-$s$ homogeneous polynomial in the oscillators
$\alpha^a$, $\alpha^z$, while the ket-vector $|\phi_{s'}\rangle$ is
degree-$s'$ homogeneous polynomial in the oscillators $\alpha^a$, i.e., these
ket-vectors satisfy the relations%
\footnote{ Throughout this paper we use the following notation for operators
constructed out the oscillators and derivatives: $N_\alpha \equiv \alpha^a
\bar\alpha^a$, $N_z \equiv \alpha^z \bar\alpha^z$, $\alpha^2 =\alpha^a
\alpha^a$, $\bar\alpha^2 = \bar\alpha^a \bar\alpha^a$,
$\Box=\partial^a\partial^a$, $\alpha\partial =\alpha^a\partial^a$,
$\bar\alpha\partial =\bar\alpha^a\partial^a$.}
\be
\label{18052008-06} (N_\alpha + N_z - s)|\phi\rangle = 0 \,,
\qquad
(N_\alpha - s')|\phi_{s'}\rangle = 0 \,.
\ee
In terms of the ket-vector $\phik$, double-tracelessness
constraint \rf{18052008-03} takes the form%
\footnote{ We adapt the formulation in terms of the double-traceless gauge
fields \cite{Fronsdal:1978vb}. Adaptation of approach in
Ref.\cite{Fronsdal:1978vb} to massive fields may be found in
Refs.\cite{Zinoviev:2001dt,Metsaev:2006zy}. Discussion of various
formulations in terms of unconstrained gauge fields may be found in
Refs.\cite{Francia:2002aa}-\cite{Buchbinder:2007ak}. Study of other
interesting approaches which seem to be most suitable for the theory of
interacting fields may be found e.g. in
Refs.\cite{Alkalaev:2003qv}-\cite{Iazeolla:2008ix}.}
\be
\label{phidoutracona01} (\bar{\alpha}^2)^2 \phik = 0 \,.
\ee

Action and Lagrangian we found take the form
\be \label{spi2lag01}  S = \int d^dx dz \ \LL\,,\qquad  \LL = \frac{1}{2}
\phibr E \phik\,, \ee
$\langle\phi| \equiv (\phik)^\dagger$, where operator $E$ is given by
\beq
\label{Frosecordope01} E  & = & E_\smtwo + E_\smone + E_\smzero\,,
\\[3pt]
\label{Frosecordope02} && E_\smtwo \equiv \Box -
\alpha\partial\bar\alpha\partial + \frac{1}{2}(\alpha\partial)^2\bar\alpha^2
+ \frac{1}{2} \alpha^2 (\bar\alpha\partial)^2 - \frac{1}{2}\alpha^2 \Box
\bar\alpha^2
-\frac{1}{4}\alpha^2\alpha\partial\,\bar\alpha\partial\bar\alpha^2\,,\qquad
\\[3pt]
&& E_\smone \equiv  \eb_1 \AA + e_1 \bar\AA \,,
\\[3pt]
&& E_\smzero \equiv m_1 + \alpha^2\bar\alpha^2m_2 + \mb_3 \alpha^2 + m_3
\bar\alpha^2\,,
\\[3pt]
&&\hspace{1cm} \AA \equiv  \alpar - \alpha^2 \albpar +
\frac{1}{4}\alpha^2\,\alpar\,\bar\alpha^2 \,,
\\[3pt]
\label{Abdef02} && \hspace{1cm} \bar\AA \equiv \albpar -  \alpar\bar\alpha^2
+ \frac{1}{4}\alpha^2\,\albpar\,\bar\alpha^2 \,,
\eeq

\beq
\label{e1def01}&& e_1 = e_{1,1} \Bigl( \partial_z + \frac{2s + d -5
-2N_z}{2z}\Bigr)\,,
\\
\label{e1def02} && \eb_1 = \Bigl(\partial_z - \frac{2s + d -5
-2N_z}{2z}\Bigr) \eb_{1,1}\,,
\\[5pt]
\label{e11def01} && e_{1,1} = \alpha^z f \,,\qquad \eb_{1,1} = f
\bar\alpha^z\,,
\\[5pt]
\label{fdef01} && f \equiv \varepsilon
\Bigl(\frac{2s+d-4-N_z}{2s+d-4-2N_z}\Bigr)^{1/2}\,, \qquad \varepsilon = \pm
1\,,
\eeq

\beq
m_1 &  =  &  \eb_1 e_1 - 2\frac{2s + d-3 -2N_z}{2s + d-4 - 2N_z} e_1 \eb_1\,,
\\[3pt]
m_2 &  =  &  - \half \eb_1 e_1 +  \frac{1}{4} \frac{2s + d -2N_z}{2s + d-4 -
2N_z} e_1 \eb_1\,,
\\[3pt]
m_3 & = & \frac{1}{2}e_1 e_1 \,,
\qquad \qquad
\mb_3  =  \frac{1}{2} \eb_1 \eb_1 \,,
\eeq
and subscript $n$ in $E_{\smx{n}}$ \rf{Frosecordope01} tells us that
$E_{\smx{n}}$ is degree-$n$ homogeneous polynomial in the flat derivative
$\partial^a$. We note that gauge invariance requires $\varepsilon^2=1$.
Because $\varepsilon$ depends on $N_z$, this leaves two possibilities
$\varepsilon =\pm 1$ at least.

The following remarks are in order.
\\
i) Operator $E_\smtwo$ \rf{Frosecordope02} is the symmetrized Fronsdal
operator represented in terms of the oscillators. This operator does not
depend on the radial coordinate and derivative, $z$, $\partial_z$, and it
takes the same form as the one of massless field in $d$-dimensional flat
space.
\\
ii) Dependence of operator $E$ \rf{Frosecordope01} on the radial coordinate
and derivative, $z$, $\partial_z$, is entirely governed by the operators
$e_1$ and $\eb_1$ which are similar to ladder operators appearing in quantum
mechanics. Sometimes, we
refer to the operators $e_1$ and $\eb_1$ as ladder operators%
\footnote{ Interesting application of other ladder operators to studying
AdS/QCD correspondence may be found in \cite{Brodsky:2008pg}. We believe that
our approach will also be useful for better understanding of various aspects
of AdS/QCD correspondence which are discussed e.g. in
\cite{Brodsky:2008pg}-\cite{Grigoryan:2007vg}.}.
\\
iii) Representation for the Lagrangian in \rf{spi2lag01} -\rf{Abdef02} is
universal and is valid for arbitrary Poincar\'e invariant theory. Various
Poincar\'e invariant theories are distinguished by ladder operators entering
the operator $E$. This is to say that the operators $E$ of massive and
conformal fields in flat space depend on the oscillators $\alpha^a$,
$\bar\alpha^a$ and the flat derivative $\partial^a$ in the same way as the
operator $E$ of $AdS$ fields \rf{Frosecordope01}. In other words, the
operators $E$ for massless $AdS$ fields, massive and conformal fields in flat
space are distinguished only by the operators $e_1$ and $\eb_1$. For example,
all that is required to get the operator $E$ for massive spin-$s$ field in
$d$-dimensional flat space is to make the substitutions
\be \label{20082008-04}
e_1 \rightarrow  m \alpha^z f\,,
\qquad
\eb_1\rightarrow  - m f\bar\alpha^z\,,
\ee
where $m$ is mass parameter of the massive field and $f$ is given in
\rf{fdef01}. Note also that our field content \rf{18052008-02} is similar to
the one of Stueckelberg formulation of massive field in $d$-dimensional space
\cite{Zinoviev:2001dt}. Expressions for $e_1$, $\eb_1$ appropriate for
conformal fields may be found in Refs.\cite{Metsaev:2007fq}.

{\it Gauge symmetries}. We now discuss gauge symmetries of Lagrangian in
\rf{spi2lag01}. To this end we introduce the following set of gauge
transformation parameters:
\be \label{epsilonset01}
\xi_{s'}^{a_1\ldots a_{s'}}\,,\qquad\qquad s'=0,1,\ldots,s-1. \ee
The gauge parameters $\xi_0$, $\xi_1^a$, and $\xi_{s'}^{a_1\ldots a_{s'}}$,
$s'\geq 2$ in \rf{epsilonset01}, are the respective scalar, vector, and
rank-$s'$ totally symmetric tensor fields of the $so(d-1,1)$ algebra. The
gauge parameters $\xi_{s'}^{a_1\ldots a_{s'}}$ with $s'\geq 2 $ are subjected
to the tracelessness constraint
\be \label{epsdoutracon01} \xi_{s'}^{aaa_3\ldots a_{s'}}=0\,, \qquad s'\geq
2\,. \ee
We now, as usually, collect gauge transformation parameters in ket-vector
$\xik$ defined by
\beq
&& |\xi\rangle \equiv \sum_{s'=0}^{s-1} (\alpha^z)^{s-1-s'}|\xi_{s'}\rangle
\,,
\\
&& |\xi_{s'}\rangle \equiv
\frac{1}{s'!\sqrt{(s -1 - s')!}}\,
\alpha^{a_1} \ldots \alpha^{a_{s'}}\, \xi_{s'}^{a_1\ldots a_{s'}}
|0\rangle\,.
\eeq
The ket-vectors $\xik$, $|\xi_{s'}\rangle$ satisfy the algebraic constraints
\be
(N_\alpha + N_z - s +1 ) \xik=0 \,,
\qquad
(N_\alpha - s' ) |\xi_{s'}\rangle=0 \,,
\ee
which tell us that $\xik$ is a degree-$(s-1)$ homogeneous polynomial in the
oscillators $\alpha^a$, $\alpha^z$, while $|\xi_{s'}\rangle$ is degree-$s'$
homogeneous polynomial in the oscillators $\alpha^a$.
In terms of the ket-vector $\xik$, tracelessness constraint
\rf{epsdoutracon01} takes the form
\be \label{xialgcon01} \bar\alpha^2 \xik=0 \,.\ee
Gauge transformation can entirely be written in terms of $\phik$ and $\xik$.
We find the following gauge transformation:
\be \label{gautraarbspi01}
\delta \phik  =  ( \alpar - e_1 - \frac{1}{2s + d- 6 -2N_z}\alpha^2\eb_1 )
\xik \,,
\ee
where $e_1$, $\eb_1$ are given in \rf{e1def01},\rf{e1def02}. From
\rf{gautraarbspi01}, we see that the flat derivative $\partial^a$ enters only
in $\alpar$-term in \rf{gautraarbspi01}, while the radial coordinate and
derivative, $z$, $\partial_z$, enter only in the operators $e_1$, $\eb_1$.
Thus, all radial coordinate and derivative contributions to gauge
transformation \rf{gautraarbspi01} are entirely expressed in terms of the
ladder operators $e_1$ and $\eb_1$%
\footnote{ Making substitutions \rf{20082008-04} in \rf{Frosecordope01} and
\rf{gautraarbspi01} one can make sure that our Lagrangian and gauge
transformation match with those of flat limit of $AdS$ massive field theory
in Ref.\cite{Zinoviev:2001dt}.}.

We finish this Section with the following remark. Introducing new mass-like
operator
\be \label{MMdef01} \MM^2 \equiv - \eb_1 e_1 + \frac{2s + d - 2 - 2N_z}{2s +
d - 4 - 2N_z} e_1 \eb_1\,, \ee
and using explicit expressions for operators $e_1$ and $\eb_1$
\rf{e1def01},\rf{e1def02} we find
\be
\label{MMdef02} \MM^2 = -  \partial_z^2 + \frac{1}{z^2} (\nu^2 -\frac{1}{4}
)\,,
\qquad\quad
\nu \equiv s + \frac{d-4}{2} - N_z\,.
\ee
We make sure that the operators $\MM^2$, $e_1$ $\eb_1$ satisfy the following
commutators:
\be
\label{e1m2com} [e_1,\MM^2] = 0\,, \qquad [\eb_1,\MM^2]= 0 \,.\ee
Because the operators $e_1$, $\eb_1$ enter gauge transformation
\rf{gautraarbspi01}, relations \rf{e1m2com} can be considered as requirement
for gauge invariance of the operator $\MM^2$. Therefore, $\MM^2$ in
\rf{MMdef01} can be considered as a definition of gauge invariant mass
operator. We note that making substitutions \rf{20082008-04} in \rf{MMdef01}
gives $\MM^2=m^2$. Thus, we see that our definition of mass operator $\MM^2$
\rf{MMdef01} gives desired result for massive field in flat space and
provides interesting generalization of notion of mass operator to the case of
massless $AdS$ field \rf{MMdef02}.

\newsection{Global $so(d,2)$ symmetries }

Relativistic symmetries of $AdS_{d+1}$ space are described by the $so(d,2)$
algebra. In our approach, the massless spin-$s$ $AdS_{d+1}$ field is
described by the set of the $so(d-1,1)$ algebra fields \rf{18052008-02}.
Therefore it is reasonable to represent the $so(d,2)$ algebra so that to
respect manifest $so(d-1,1)$ symmetries. For application to the AdS/CFT
correspondence, most convenient form of the $so(d,2)$ algebra that respects
the manifest $so(d-1,1)$ symmetries is provided by nomenclature of the
conformal algebra. This is to say that the $so(d,2)$ algebra consists of
translation generators $P^a$, conformal boost generators $K^a$, dilatation
generator $D$, and generators $J^{ab}$ which span $so(d-1,1)$ algebra. We use
the following normalization for commutators of the $so(d,2)$ algebra
generators:
\beq
\label{ppkk}
&& {}[D,P^a]=-P^a\,, \hspace{2cm}  {}[P^a,J^{bc}]=\eta^{ab}P^c -\eta^{ac}P^b
\,,
\\
&& [D,K^a]=K^a\,, \hspace{2.2cm} [K^a,J^{bc}]=\eta^{ab}K^c - \eta^{ac}K^b\,,
\\[5pt]
\label{pkjj} && \hspace{2.5cm} {}[P^a,K^b]=\eta^{ab}D - J^{ab}\,,
\\
&& \hspace{2.5cm} [J^{ab},J^{ce}]=\eta^{bc}J^{ae}+3\hbox{ terms} \,.
\eeq
Requiring $so(d,2)$ symmetries implies that the action is invariant with
respect to transformation $\delta_{\hat{G}} \phik  = \hat{G} \phik$, where
the realization of $so(d,2)$ algebra generators $\hat{G}$ in terms of
differential operators takes the form
\beq
\label{conalggenlis01} && P^a = \partial^a \,,
\qquad
J^{ab} = x^a\partial^b -  x^b\partial^a + M^{ab}\,,
\\[3pt]
\label{conalggenlis03} && D = x\partial  + \Delta\,,
\qquad
\Delta \equiv  z\partial_z + \frac{d-1}{2}\,,
\\[3pt]
\label{conalggenlis04} && K^a = -\frac{1}{2}x^2\partial^a + x^a D + M^{ab}x^b
+ R^a \,,
\eeq
$x\partial\equiv x^a\partial^a$, $x^2\equiv x^ax^a$. In
\rf{conalggenlis01},\rf{conalggenlis04}, $M^{ab}$ is spin operator of the
$so(d-1,1)$ algebra. Commutation relations for $M^{ab}$ and representation of
$M^{ab}$ on space of ket-vector $\phik$ \rf{18052008-04} take the form
\be  [M^{ab},M^{ce}]=\eta^{bc}M^{ae}+3\hbox{ terms} \,, \qquad
M^{ab} = \alpha^a \bar\alpha^b - \alpha^b \bar\alpha^a\,. \ee
Operator $R^a$ appearing in $K^a$ \rf{conalggenlis04} is given by
\beq
R^a & = &  -z \Cwt^a \eb_{1,1} + z e_{1,1} \bar\alpha^a - \half z^2
\partial^a\,,
\\[1pt]
&& \ \ \ \ \ \Cwt^a \equiv \alpha^a - \alpha^2 \frac{1}{2N_\alpha + d
-2}\bar\alpha^a\,,
\eeq
where $e_{1,1}$, $\eb_{1,1}$ are given in \rf{e11def01}. We see that
realization of Poincar\'e symmetries on bulk $AdS$ fields \rf{conalggenlis01}
coincide with realization of Poincar\'e symmetries on boundary CFT operators.
Note that realization of $D$- and $K^a$-symmetries on bulk $AdS$ fields
\rf{conalggenlis03},\rf{conalggenlis04} coincides, by module of contributions
of operators $\Delta$ and $R^a$, with the realization of $D$- and
$K^a$-symmetries on boundary CFT operators. Realizations of the $so(d,2)$
algebra on bulk $AdS$ fields and boundary CFT operators are distinguished by
$\Delta$ and $R^a$. The realization of the $so(d,2)$ symmetries on bulk $AdS$
fields given in \rf{conalggenlis01}-\rf{conalggenlis04} turns out to be very
convenient for studying AdS/CFT correspondence \cite{Metsaev:2008fs}.

\newsection{  Modified de Donder gauge}

We begin with discussion of gauge-fixing procedure at the level of Lagrangian
\rf{spi2lag01}. We find that use of the following {\it modified de Donder}
gauge-fixing term
\beq
\label{22082008-01} \LL_{g.fix} & = & \half \phibr E_{g.fix} \phik \,,
\qquad \ \ \ \
E_{g.fix} \equiv  C\bar{C}  \,,
\\[3pt]
\label{10052008-02} && C \equiv \alpar - \half \alpha^2 \albpar - e_1
\Pi^\smponetwo + \half \eb_1 \alpha^2\,,
\\[3pt]
\label{10052008-01} && \bar{C} \equiv  \albpar - \half \alpar \bar\alpha^2 +
\half e_1 \bar\alpha^2 - \eb_1\Pi^\smponetwo\,,
\\[1pt]
\label{18052008-08} && \Pi^\smponetwo \equiv 1 -\alpha^2\frac{1}{2(2N_\alpha
+d)}\bar\alpha^2\,,
\eeq
leads to the surprisingly simple gauge fixed Lagrangian $\LL_{total}$:
\beq \label{20082008-05}
&& \LL_{total} \equiv \LL + \LL_{g.fix} \,,
\\
\label{20082008-06} && \LL_{total} = \half \phibr E_{total} \phik \,,
\\
\label{20082008-07} && E_{total} =  (1-\frac{1}{4}\alpha^2 \bar\alpha^2)
(\Box - \MM^2)\,,
\eeq
where $\MM^2$ is given in \rf{MMdef02}.%
\footnote{ Making substitutions \rf{20082008-04} in
\rf{10052008-02},\rf{10052008-01} gives gauge fixed Lagrangian for massive
field of the form \rf{20082008-05}-\rf{20082008-07} with $\MM^2 = m^2$.}
We note that our gauge-fixing term \rf{22082008-01} respects the Poincar\'e
and dilatation symmetries but breaks the conformal boost $K^a$-symmetries,
i.e., the simple form of gauge fixed Lagrangian \rf{20082008-05} is achieved
at the cost of the $K^a$-symmetries.

We now discuss gauge-fixing procedure at the level of equations of motion. To
this end we note that gauge invariant Lagrangian \rf{spi2lag01}
leads to the following equations of motion:%
\footnote{ Appearance of the projector $\Pi^\smptwothree$ in equations of
motion \rf{eqsmot01} is related to the fact that the operators $E_\smone$,
$E_\smzero$, in contrast to the symmetrized Fronsdal operator $E_\smtwo$, do
not respect double tracelessness constraint \rf{phidoutracona01}. Note that
the ket-vectors $E_\smone\phik$, $E_\smzero\phik$ are triple-traceless,
$(\bar\alpha^2)^3E_\smone\phik=0$, $(\bar\alpha^2)^3 E_\smzero\phik=0$.}
\beq \label{eqsmot01}
&& \Bigl( E_\smtwo + \Pi^\smptwothree (E_\smone + E_\smzero)\Bigr) \phik=
0\,,
\\
&& \qquad \quad \ \Pi^\smptwothree \equiv 1 - (\alpha^2)^2
\frac{1}{8(2N_\alpha +d)(2N_\alpha +d+2)}(\bar\alpha^2)^2\,.
\eeq
These equations can be represented as
\be \label{eqsmot02} (\EE_\smtwo + \EE_\smone + \EE_\smzero ) \phik = 0\,,\ee
\beq
\EE_\smtwo & \equiv & \Box - \alpar\albpar + \half (\alpar)^2 \bar\alpha^2\,,
\\[1pt]
\EE_\smone & \equiv & e_1(\albpar - \alpar \bar\alpha^2)
\nonumber\\[1pt]
& + & \eb_1 \Bigl(\alpar + \alpha^2 \frac{1}{2N_\alpha + d-2} \albpar -
\alpha^2 \alpar \frac{1}{2N_\alpha + d} \bar\alpha^2\Bigr)\,,
\\[1pt]
\EE_\smzero & = &  m_1 - (m_1 + 4m_2)\alpha^2 \frac{1}{2(2N_\alpha +
d-2)}\bar\alpha^2 + m_3 \bar\alpha^2 - \mb_3 \alpha^2 \frac{2}{2N_\alpha +
d-2} \Pi^\smponetwo \,. \ \ \ \
\eeq
{\it Modified de Donder gauge} condition is then defined to be
\be \label{080405-01} \bar{C} \phik = 0\,, \ee
where the operator $\bar{C}$ is given in \rf{10052008-01}. Because of
double-tracelessness of $\phik$ \rf{phidoutracona01}, operator $\bar{C}$
\rf{10052008-01} satisfies the relation $\bar\alpha^2\bar{C}\phik=0$, i.e.,
gauge condition \rf{080405-01} respects constraint for gauge transformation
parameter $\xik$, \rf{xialgcon01}. Using the modified de Donder gauge
condition in gauge invariant equations of motion \rf{eqsmot02} leads to the
following gauge fixed equations of motion:
\be \label{20082008-03} (\Box - \MM^2 )\phik = 0 \,, \ee
where $\MM^2$ is defined in \rf{MMdef02}. In terms of fields
\rf{18052008-02}, equation \rf{20082008-03} can be represented as
\be \label{23082008-01} \Bigl( \Box + \partial_z^2 - \frac{1}{z^2}(\nu_{s'}^2
- \frac{1}{4})\Bigr)\phi_{s'}^{a_1\ldots a_{s'}} = 0\,,
\qquad\quad
\nu_{s'} \equiv  s' + \frac{d-4}{2}\,, \ee
$s'=0,1,\ldots, s$. Thus, our {\it modified de Donder gauge condition
\rf{080405-01} leads to decoupled equations of motion} \rf{23082008-01} which
can easily be solved in terms of the Bessel function
\footnote{ Interesting method of solving $AdS$ field equations of motion
which is based on star algebra products in auxiliary spinor variables is
discussed in Ref.\cite{Bolotin:1999fa}.}.
For spin-1 field, gauge condition \rf{080405-01}, found in
\cite{Metsaev:1999ui}, turns out to be a modification of the Lorentz gauge.

We note that equations of motion \rf{20082008-03} have on-shell leftover
gauge symmetries. These on-shell leftover gauge symmetries can simply be
obtained from generic gauge symmetries \rf{gautraarbspi01} by the
substituting $\xik \rightarrow |\xi_{lfov}\rangle$, where the
$|\xi_{lfov}\rangle$ satisfies the following equations of motion:
\be (\Box - \MM^2 )|\xi_{lfov}\rangle = 0 \,. \ee

\newsection{ Comparison of standard and modified de Donder gauges}

Our approach to the massless spin-$s$ field in $AdS_{d+1}$ is based on use of
double-traceless $so(d-1,1)$ algebra fields \rf{18052008-02}. One of popular
approaches to the massless spin-$s$ field in $AdS_{d+1}$ is based on use of
double-traceless $so(d,1)$ algebra field $\Phi^{A_1\ldots A_s}$
\cite{Fronsdal:1978vb}. The aim of this Section is twofold. First we explain
how our modified de Donder gauge is represented in terms of the commonly used
field $\Phi^{A_1\ldots A_s}$. Also, we compare
the modified de Donder gauge and commonly used standard de Donder gauge%
\footnote{ Recent applications of the {\it standard} de Donder gauge to the
various problems of higher-spin fields may be found in
Refs.\cite{Guttenberg:2008qe,Manvelyan:2008ks}.}.
Second we show explicitly how our fields
\rf{18052008-02} are related to the field $\Phi^{A_1\ldots A_s}$.

We begin with discussion of modified de Donder gauge-fixing procedure at the
level of Lagrangian. First we present gauge invariant Lagrangian for the
field $\Phi^{A_1\ldots A_s}$. To simplify presentation we introduce, as
before, the following ket-vector
\beq
\label{froketvect} && |\Phi\rangle = \frac{1}{s!}\Phi^{A_1\ldots A_s}
\alpha^{A_1}\ldots \alpha^{A_s} |0\rangle\,,
\\[3pt]
\label{froketvectcon} && (\bar\alphabf^2)^2 |\Phi\rangle=0,
\\[3pt]
\label{alphabfdef01} && \alphabf^2 \equiv \alpha^A\alpha^A \,,\qquad
\bar\alphabf^2 \equiv \bar\alpha^A\bar\alpha^A\,,
\eeq
where \rf{froketvectcon} tells us that the $\Phi^{A_1\ldots A_s}$ is
double-traceless, and the scalar products like $\alpha^A\alpha^A$ are
decomposed as $\alpha^A\alpha^A = \alpha^a\alpha^a +\alpha^z\alpha^z$. In
terms of $\Phik$, gauge invariant Lagrangian
takes the form%
\footnote{ Since Ref.\cite{Fronsdal:1978vb}, various approaches to massless
totally symmetric $AdS$ fields were developed in the literature (see e.g.
\cite{Lopatin:1987hz,Metsaev:1999ui,Buchbinder:2001bs,Hallowell:2005np}). We
use setup discussed in Ref.\cite{Metsaev:1999ui}. Formulas in
Ref.\cite{Fronsdal:1978vb} are adapted to $AdS_4$ with mostly negative metric
tensor, while our formulas are adapted to $AdS_{d+1}$ with mostly positive
metric tensor. Taking this into account and plugging $d=3$ in
\rf{lagstand01add01} we make sure that our operator $\EE$ matches with the
operator $L_0$ in Eq.(2.7) in Ref.\cite{Fronsdal:1978vb}.}
\beq
\label{lagstand01} && \LL = \half e \Phibr
(1-\frac{1}{4}\alphabf^2\bar\alphabf^2) \EE \Phik \,,
\\[1pt]
\label{lagstand01add01} && \EE \equiv  \Box_{AdS} - \alphabf {\bf
D}\bar\alphabf {\bf D} +\frac{1}{2}(\alphabf{\bf  D})^2\bar\alphabf^2 -
s(s+d-5) + 2d-4 -\alphabf^2\bar\alphabf^2\,,
\\[1pt]
&& \Box_{AdS}\equiv D^A D^A + \omega^{AAB}D^B\,,
\qquad
\bar\alphabf {\bf D} \equiv \bar\alpha^A D^A\,,
\qquad
\alphabf {\bf D} \equiv \alpha^A D^A\,,
\eeq
where $e=\det e_\mu^A$, $e_\mu^A$ stands for vielbein of $AdS_{d+1}$ space,
and $D^A$ are covariant derivatives (for details of notation, see Appendix).
Lagrangian \rf{lagstand01} can be represented as
\beq
\label{24082008-01} \LL & = & \half e \Phibr E \Phik \,,
\\
E & \equiv & \Box_{AdS} - \alphabf \Dbf \bar\alphabf \Dbf +
\frac{1}{2}(\alphabf \Dbf)^2\bar\alphabf^2 + \frac{1}{2} \alphabf^2
(\bar\alphabf \Dbf)^2 - \frac{1}{2}\alphabf^2 \Box_{AdS} \bar\alphabf^2
-\frac{1}{4}\alphabf^2\alphabf \Dbf\,\bar\alphabf \Dbf \bar\alphabf^2
\nonumber\\
& - & s(s+d-5) + 2d-4 + \half(s(s+d-3) -d)\alphabf^2\bar\alphabf^2\,.
\eeq

We now ready to discuss the modified de Donder gauge. To make our study more
useful we discuss both the modified and standard de Donder gauges. Note that
our formulas for standard de Donder gauge are valid for arbitrary
parametrization of $AdS$, while the ones for modified de Donder gauge are
adapted to the Poincar\'e parametrization. Gauge-fixing term is defined to be
\be \LL_{g.fix} = \half e \Phibr E_{g.fix} \Phik \,, \ee
where operator $E_{g.fix}$ corresponding to the standard de Donder gauge
fixing and the modified de Donder gauge fixing is given by
\be
E_{g.fix} =  \left\{ \begin{array}{ll}
\Cbf_{stand}  \bar{\Cbf}_{stand} \,, & \hbox{ standard gauge};
\\[7pt]
\Cbf_{mod} \bar{\Cbf}_{mod}  \,, & \hbox{ modified gauge},
\end{array}\right.
\ee
and we use the notation
\beq
\label{defgaucon001} && \Cbf_{stand} \equiv  \alphabf {\bf D} - \half
\alphabf^2 \bar\alphabf {\bf D} \,,
\qquad \ \ \ \ \
\bar{\Cbf}_{stand} \equiv \bar\alphabf {\bf D} - \half \alphabf {\bf D}
\bar\alphabf^2\,,
\\[3pt]
\label{defgaucon003} &&  \Cbf_{mod} \equiv \Cbf_{stand} - 2 \Cbf_\perp^z \,,
\hspace{1.8cm}
\bar{\Cbf}_{mod} \equiv \bar{\Cbf}_{stand} + 2\bar{\Cbf}_\perp^z \,,
\\[5pt]
\label{defgaucon005} && \Cbf_\perp^z \equiv \alpha^z  - \half \alphabf^2
\bar\alpha^z \,,
\hspace{2.5cm}
\bar{\Cbf}_\perp^z \equiv \bar \alpha^z -\half \alpha^z \bar\alphabf^2 \,.
\eeq
We now make sure that the gauge fixed Lagrangian $\LL_{total}$ takes the form
\beq
&& \LL_{total}  \equiv \LL + \LL_{g.fix} \,,
\\[3pt]
&& \LL_{total} = \half e
\Phibr(1-\frac{1}{4}\alphabf^2\bar\alphabf^2)\EE_{total} \Phik \,,
\eeq
\be\label{eetot03}
\EE_{total} =  \left\{ \begin{array}{ll}
\Box_{AdS} -s(s+d-5) +2d - 4 - \alphabf^2\bar\alphabf^2 \,, & \hbox{ standard
gauge};
\\[5pt]
\!\!\Box_{0\,AdS} - s(s+d-4) + 2d-4 - \alphabf^2\bar{\alpha}_z^2+ (2s + d -
5)N_z \,, & \hbox{ modified gauge},
\end{array}\right.
\ee
where $\Box_{0\,AdS} \equiv z^2(\Box+\partial_z^2) +(1-d)z\partial_z$.
Alternatively, the operator $\EE_{total}$ corresponding to the modified de
Donder gauge in \rf{eetot03} can be represented as
\be
\EE_{total} = \Box_{0\,AdS} - \alpha^2 \bar{\alpha}^z\bar{\alpha}^z - \nu^2 +
\frac{d^2}{4} \,,
\ee
where $\nu$ is given in \rf{MMdef02}.

We proceed with discussion of gauge-fixing procedure at the level of
equations of motion. To this end we note that gauge invariant Lagrangian
\rf{lagstand01} leads to the following equations of motion:
\be \label{28082008-01} \EE \Phik=0 \,.\ee
We now define the standard and modified de Donder gauge conditions as
\beq
\label{20082008-01xx} && \bar{\Cbf}_{stand}|\Phi\rangle  =  0  \,,
\hspace{1cm} \hbox{ standard de Donder gauge};
\\[5pt]
\label{20082008-01} && \bar{\Cbf}_{mod}|\Phi\rangle  =  0 \,, \hspace{1.3cm}
\hbox{ modified de Donder gauge},
\eeq
where $\bar{\Cbf}_{stand}$, $\bar{\Cbf}_{mod}$ are given in
\rf{defgaucon001},\rf{defgaucon003}. Using
\rf{20082008-01xx},\rf{20082008-01} in \rf{28082008-01} we get gauge fixed
equations of motion
\be \label{20082008-02} \EE_{total} \Phik=0 \,,\ee
where $\EE_{total}$ is given in \rf{eetot03}. We note that, because of
$\Cbf_\perp^z$- and $\bar{\Cbf}_\perp^z$-terms, the modified de Donder gauge
breaks some of the $so(d,2)$ symmetries. In the conformal algebra
nomenclature, these broken symmetries correspond to broken conformal boost
$K^a$-symmetries.

From $\EE_{total}$ \rf{eetot03}, we see that, because of
$\alphabf^2\bar{\alpha}^z\bar{\alpha}^z$-term, the modified de Donder gauge
for $\Phik$ does not lead to decoupled equations for
the ket-vector $\Phik$ when%
\footnote{ For spin-1 field, gauge condition \rf{20082008-01} and the
corresponding decoupled equations of motion were found in
\cite{Metsaev:1999ui}.}
$s\geq 2$. It turns out that in order to obtain decoupled equations of motion
we should introduce our set of fields in \rf{18052008-02}. We remind that
$\Phik$ is a double-traceless field \rf{froketvectcon} of the $so(d,1)$
algebra, while $\phik$ describes double-traceless fields \rf{phidoutracona01}
of the $so(d-1,1)$ algebra. This is to say that to get decoupled equations of
motion we have to make transformation from the $so(d,1)$ ket-vector $\Phik$
to $so(d-1,1)$ ket-vector $\phik$. We find the following transformation from
the ket-vector $\Phik$ to our ket-vector $\phik$:
\be \label{05012008} \phik  =  z^{\frac{1-d}{2}}\NN  \Pi^{\phi\Phi}\Phik
\,,\ee
\beq \label{05012008-05-01}
\Pi^{\phi\Phi}
& \equiv & \Pi_\alpha^\smpone + \alpha^2 \frac{1}{2(2N_\alpha + d)}
\Pi_\alpha^\smpone (\bar\alpha^2 + \frac{2N_\alpha +d}{2N_\alpha + d
-2}\bar\alpha^z\bar\alpha^z)\,,
\\[3pt]
\label{05012008-05-02} && \Pi_\alpha^\smpone \equiv
\Pi^\smpone(\alpha,0,N_\alpha,\bar\alpha,0,d)\,,
\\[3pt]
\label{05012008-05-03} && \NN \equiv \Bigl(\frac{2^{N_z} \Gamma( N_\alpha +
N_z + \frac{d-3}{2}) \Gamma( 2N_\alpha + d - 3)}{\Gamma(N_\alpha +
\frac{d-3}{2})\Gamma(2N_\alpha + N_z + d - 3)}\Bigr)^{1/2}\,,
\\[3pt]
\label{05012008-05-04} && N_\alpha =\alpha^a\bar\alpha^a\,, \quad N_z
=\alpha^z\bar\alpha^z \,,
\eeq
where $\Gamma$ is Euler gamma function and operator $\Pi_\alpha^\smpone$ in
\rf{05012008-05-02} is obtained from the function
\be \label{pibasicdef01}
\Pi^\smpone(\alpha,\alpha^z, X,\bar\alpha,\bar\alpha^z, Y)
\equiv \sum_{n=0}^\infty (\alpha^2+\alpha^z\alpha^z)^n \frac{(-)^n \Gamma(X +
\frac{Y-2}{2} + n)}{4^n n! \Gamma(X + \frac{Y-2}{2} +
2n)}(\bar\alpha^2+\bar\alpha^z\bar\alpha^z)^n\,,
\ee
by equating $\alpha^z=\bar\alpha^z=0$, $X=N_\alpha$, $Y=d$. We introduce the
$z$-factor in r.h.s. of \rf{05012008} to obtain canonically normalized
ket-vector $\phik$.

Inverse transform of \rf{05012008} takes the form
\be \label{05012008-20} \Phik = z^{\frac{d-1}{2}} \Pi^{\Phi\phi} \NN \phik\,,
\ee
\beq
\Pi^{\Phi\phi}
& \equiv & \Pi_{\alphabf}^\smpone +  \alphabf^2 \frac{1}{2(2 N_\alphabf + d
+1)}\Pi_{\alphabf}^\smpone (\bar\alpha^2 - \frac{2}{2 N_\alphabf + d
-1}\bar\alpha^z\bar\alpha^z) \,,
\\[3pt]
\label{05012008-05-02bf} && \Pi_{\alphabf}^\smpone \equiv
\Pi^\smpone(\alpha,\alpha^z,N_\alphabf,\bar\alpha,\bar\alpha^z,d+1)\,,\qquad
 \ N_\alphabf \equiv N_\alpha + N_z\,,
\eeq
where $\alphabf^2$ is given in \rf{alphabfdef01} and $\Pi_{\alphabf}^\smpone$
is obtained from $\Pi^\smpone$ \rf{pibasicdef01} by equating $X=N_\alphabf$,
$Y=d+1$.

We now ready to compare modified de Donder gauges for $\phik$ \rf{080405-01}
and $\Phik$ \rf{20082008-01}. Inserting \rf{05012008-20} in \rf{20082008-01}
and choosing $\varepsilon=-1$ in \rf{fdef01}, we make sure that modified de
Donder gauge for $\Phik$ \rf{20082008-01} amounts to modified de Donder gauge
for $\phik$ \rf{080405-01} i.e., modified de Donder gauges for $\phik$
\rf{080405-01} and $\Phik$ \rf{20082008-01} match. Also we make sure that
inserting \rf{05012008-20} in equations \rf{20082008-02} leads to equations
\rf{20082008-03}, i.e., equations of motions for $\phik$ and $\Phik$ match.
Finally, one can make sure that gauge invariant Lagrangian for $\Phik$
\rf{lagstand01} and the one for $\phik$ \rf{spi2lag01} match.

We now compare gauge transformation of the ket-vector $\phik$
\rf{gautraarbspi01} and gauge transformation of $\Phik$ which takes the form
\be \label{gautraarbspi01bf} \delta \Phik = \alphabf \Dbf \Xik\,, \qquad \Xik
= \frac{1}{(s-1)!} \alpha^{A_1}\ldots \alpha^{A_{s-1}}\Xi^{A_1\ldots
A_{s-1}}|0\rangle\,,
\ee
where gauge transformation parameter $\Xi^{A_1\ldots A_{s-1}}$  is traceless,
$\Xi^{AAA_3\ldots A_{s-1}}=0$, i.e., $\bar\alphabf^2\Xik=0$. To this end we
note that gauge transformation parameters $\xik$ and $\Xik$ are related as
\be \label{22082008-02} \xik  =  z^{\frac{3-d}{2}} \NN'  \Pi_\alpha^\smpone
\Xik \,,\quad
\qquad
\Xik  =  z^{\frac{d-3}{2}}  \Pi_{\alphabf}^\smpone \NN' \xik \,,\qquad \ee
\be \NN' \equiv \NN|_{N_\alpha\rightarrow N_\alpha+1}\,, \ee
where $\Pi_\alpha^\smpone$, $\Pi_{\alphabf}^\smpone$, $\NN$ are given in
\rf{05012008-05-02},\rf{05012008-05-02bf}, \rf{05012008-05-03} respectively.
We note that $\Pi_\alpha^\smpone$ and $\Pi_{\alphabf}^\smpone$ are projectors
on traceless ket-vectors, i.e., if $|\phi_{trf}\rangle$ and
$|\Phi_{trf}\rangle$ are tracefull ket-vectors, $\bar\alpha^2
|\phi_{trf}\rangle \ne 0$, $\bar\alphabf^2 |\Phi_{trf}\rangle \ne 0$, then on
has the relations $\bar\alpha^2\Pi_\alpha^\smpone|\phi_{trf}\rangle = 0$,
$\bar\alphabf^2\Pi_\alphabf^\smpone|\Phi_{trf}\rangle = 0$. Using
\rf{05012008-20},\rf{22082008-02}, and $\varepsilon=-1$ in \rf{fdef01}, we
make sure that gauge transformations \rf{gautraarbspi01} and
\rf{gautraarbspi01bf} match.

Finally we compare realization of $so(d,2)$ symmetries on the ket-vectors
$\phik$ and $\Phik$. To this end we note that on space of $\Phik$ realization
of the $so(d,2)$ algebra transformations takes the form
\beq
\label{conalggenlis01fr} && \delta_{P^a}\Phik  = \partial^a \Phik \,,
\qquad
\ \ \ \ \ \delta_{J^{ab}}\Phik = (x^a\partial^b -  x^b\partial^a +
M^{ab})\Phik\,,
\\[3pt]
\label{conalggenlis03fr} && \delta_D\Phik = x^B\partial^B \Phik\,,
\
\qquad \label{conalggenlis04fr} \delta_{K^a} \Phik = (-\frac{1}{2}x^Bx^B
\partial^a + x^a x^B\partial^B + M^{aB}x^B)\Phik \,,
\eeq
where $x^Bx^B=x^bx^b + z^2$, $x^B\partial^B = x^b\partial^b + z\partial_z$,
$M^{aB}x^B = M^{ab}x^b - M^{za}z$. Comparing \rf{conalggenlis01} and
\rf{conalggenlis01fr}, we see that the realizations of Poincar\'e symmetries
on $\phik$ and $\Phik$ match from the very beginning. Taking into account
$z$-factor in \rf{05012008-20}, it is easily seen that $D$-transformations
for $\phik$ \rf{conalggenlis03} and $\Phik$ \rf{conalggenlis03fr} also match.
All that remains to do is to match conformal boost $K^a$-transformations
given in \rf{conalggenlis04} and \rf{conalggenlis04fr}. Choosing
$\varepsilon=-1$ in \rf{fdef01}, we make sure that realizations of the
operator $K^a$ on $\phik$ \rf{conalggenlis04} and on $\Phik$
\rf{conalggenlis04fr} match.

To summarize, using the Poincar\'e parametrization of $AdS$ space, we have
developed the CFT adapted formulation of massless arbitrary spin $AdS$ field.
In our approach, Poincar\'e symmetries of the Lagrangian are manifest. As is
well known string theory solutions like $AdS_{d+1}\times S^{d+1}$ and
$Dp$-brane backgrounds supported by RR-charges have the respective the $d$-
and $(p+1)$-dimensional Poincar\'e symmetries. We note that the structure of
the Lagrangian we obtained for $AdS$ field is valid for any theory that
respects Poincar\'e symmetries. Various theories are distinguished by
appropriate ladder operators. Therefore we think that our approach might be a
good starting point for formulation of higher-spin gauge fields theory in
$AdS_{d+1}\times S^{d+1}$ and $Dp$-brane backgrounds. For the case of
$AdS_{d+1}$ field, the ladder operators depend on the radial coordinate and
the radial derivative. It would be interesting to unravel a structure and
role of ladder operators in $AdS_{d+1}\times S^{d+1}$ and
$Dp$-brane backgrounds%
\footnote{ It would also be interesting to unravel the ladder operators in
the tensionless limit of $AdS$ strings \cite{Tseytlin:2002gz,Bonelli:2003zu}.
Also we think that formalism developed in this paper might be useful for the
study of $(A)dS$ massive fields \cite{Zinoviev:2001dt} and $(A)dS$
partial-massless fields \cite{Deser:1983mm}-\cite{Skvortsov:2006at}.}.
The $AdS_{d+1}\times S^{d+1}$ and $Dp$-brane backgrounds play important role
in studying string/gauge theory dualities. Developing a theory of higher-spin
gauge fields in these backgrounds might be useful for better understanding
string/gauge theory dualities.

{\bf Acknowledgments}. This work was supported by the RFBR Grant
No.08-02-00963, RFBR Grant for Leading Scientific Schools, Grant No.
1615.2008.2, by the Dynasty Foundation and by the Alexander von Humboldt
Foundation Grant PHYS0167.

\setcounter{section}{0} \setcounter{subsection}{0}
\appendix{ Notation }

Vector indices of the $so(d-1,1)$ algebra  take the values $a,b,c=0,1,\ldots
,d-1$, while vector indices of the $so(d,1)$ algebra take the values
$A,B,C=0,1,\ldots ,d-1,d$. We use mostly positive flat metric tensors
$\eta^{ab}$, $\eta^{AB}$. To simplify our expressions we drop $\eta_{ab}$,
$\eta_{AB}$ in the respective scalar products, i.e., we use $X^a Y^a \equiv
\eta_{ab}X^a Y^b$, $X^A Y^A \equiv \eta_{AB}X^A Y^B$. Using the
identification $X^d \equiv X^z$ gives the following decomposition of the
$so(d,1)$ algebra vector: $X^A=X^a,X^z$. This implies $X^AY^A = X^aY^a +
X^zY^z$.

We use the creation operators $\alpha^a$, $\alpha^z$, and the respective
annihilation operators $\bar{\alpha}^a$, $\bar{\alpha}^z$,
\be
[\bar{\alpha}^a,\alpha^b]=\eta^{ab}\,, \qquad [\bar\alpha^z,\alpha^z]=1\,,
\qquad
\bar\alpha^a |0\rangle = 0\,,\qquad  \bar\alpha^z |0\rangle = 0\,.\ee
These operators are referred to as oscillators in this paper. The oscillators
$\alpha^a$, $\bar\alpha^a$ and $\alpha^z$, $\bar\alpha^z$, transform in the
respective vector and scalar representations of the $so(d-1,1)$ algebra and
satisfy the hermitian conjugation rules, $\alpha^{a\dagger} = \bar\alpha^a$,
$\alpha^{z\dagger} = \bar\alpha^z$. Oscillators  $\alpha^a$, $\alpha^z$ and
$\bar\alpha^a$, $\bar\alpha^z$ are collected into the respective $so(d,1)$
algebra oscillators $\alpha^A =\alpha^a,\alpha^z$ and $\bar\alpha^A
=\bar\alpha^a,\bar\alpha^z$.

$x^A = x^a,z $ denote coordinates in $d+1$-dimensional $AdS_{d+1}$ space,
\be \label{speech01}
ds^2 = \frac{1}{z^2}(dx^a dx^a + dz dz)\,,
\ee
while $\partial_A=\partial_a,\partial_z$ denote the respective derivatives,
$\partial_a \equiv \partial / \partial x^a$, $\partial_z \equiv
\partial / \partial z$. We use the notation $\Box=\partial^a\partial^a$,
$\alpha\partial =\alpha^a\partial^a$, $\bar\alpha\partial
=\bar\alpha^a\partial^a$, $\alpha^2 = \alpha^a\alpha^a$, $\bar\alpha^2 =
\bar\alpha^a\bar\alpha^a$.

The covariant derivative $D^A$ is given by $D^A = \eta^{AB}D_B$,
\be \label{vardef01}
D_A \equiv e_A^\mu D_\mu\,,  \qquad D_\mu \equiv
\partial_\mu
+\frac{1}{2}\omega_\mu^{AB}M^{AB}\,, \qquad M^{AB} \equiv \alpha^A
\bar\alpha^B - \alpha^B \bar\alpha^A\,, \ee
$\partial_\mu = \partial/\partial x^\mu$, where $e_A^\mu$ is inverse vielbein
of $AdS_{d+1}$ space, $D_\mu$ is the Lorentz covariant derivative and the
base manifold index takes values $\mu = 0,1,\ldots, d$. The $\omega_\mu^{AB}$
is the Lorentz connection of $AdS_{d+1}$ space, while $M^{AB}$ is a spin
operator of the Lorentz algebra $so(d,1)$. Note that $AdS_{d+1}$ coordinates
$x^\mu$ carrying the base manifold indices are identified with coordinates
$x^A$ carrying the flat vectors indices of the $so(d,1)$ algebra, i.e., we
assume $x^\mu = \delta_A^\mu x^A$, where $\delta_A^\mu$ is Kronecker delta
symbol. $AdS_{d+1}$ space contravariant tensor field, $\Phi^{\mu_1\ldots
\mu_s}$, is related with field carrying the flat indices, $\Phi^{A_1\ldots
A_s}$, in a standard way $\Phi^{A_1\ldots A_s} \equiv e_{\mu_1}^{A_1}\ldots
e_{\mu_s}^{A_s} \Phi^{\mu_1\ldots \mu_s}$. Helpful commutators are given by
\be   [D^A,D^B]=\Omega^{ABC} D^C - M^{AB}\,,
\qquad [\bar\alphabf \Dbf, \alphabf \Dbf]= \Box_{AdS} +
\frac{1}{2}M^{AB}M^{AB}\,,
\ee
where $\Omega^{ABC} = -\omega^{ABC}+\omega^{BAC}$ is a contorsion tensor and
we define $\omega^{ABC}\equiv e^{A\mu} \omega_\mu^{BC}$.

For the Poincar\'e parametrization of $AdS_{d+1}$ space, vielbein
$e^A=e^A_\mu dx^\mu$ and Lorentz connection, $de^A+\omega^{AB}\wedge e^B=0$,
are given by
\be\label{eomcho01} e_\mu^A=\frac{1}{z}\delta^A_\mu\,,\qquad
\omega^{AB}_\mu=\frac{1}{z}(\delta^A_z\delta^B_\mu
-\delta^B_z\delta^A_\mu)\,. \ee
With choice made in \rf{eomcho01}, the covariant derivative takes the form
$D^A= z \partial^A + M^{zA}$, $\partial^A=\eta^{AB}\partial_B$.


\small
\begin{thebibliography}{30}

\parskip=-1.pt



\bibitem{Maldacena:1997re}
  J.~M.~Maldacena,
  Adv.\ Theor.\ Math.\ Phys.\  {\bf 2}, 231 (1998)
  [Int.\ J.\ Theor.\ Phys.\  {\bf 38}, 1113 (1999)]
  [arXiv:hep-th/9711200].




\bibitem{Bekaert:2005vh}
  X.~Bekaert, S.~Cnockaert, C.~Iazeolla and M.~A.~Vasiliev,
  arXiv:hep-th/0503128.


\bibitem{Sorokin:2004ie}
  D.~Sorokin,
  AIP Conf.\ Proc.\  {\bf 767}, 172 (2005)
  [arXiv:hep-th/0405069].


\bibitem{Fotopoulos:2008ka}
  A.~Fotopoulos and M.~Tsulaia,
  arXiv:0805.1346 [hep-th].



\bibitem{Metsaev:1998it}
  R.~R.~Metsaev and A.~A.~Tseytlin,
  Nucl.\ Phys.\  B {\bf 533}, 109 (1998)
  [arXiv:hep-th/9805028].

\bibitem{Metsaev:2000ds}
  R.~R.~Metsaev,
  Class.\ Quant.\ Grav.\  {\bf 18}, 1245 (2001)
  [arXiv:hep-th/0012026].

\bibitem{Beisert:2008iq}
  N.~Beisert, R.~Ricci, A.~Tseytlin and M.~Wolf,
  arXiv:0807.3228 [hep-th].


\bibitem{Berkovits:2008ic}
  N.~Berkovits and J.~Maldacena,
  JHEP {\bf 0809}, 062 (2008)
  [arXiv:0807.3196 [hep-th]].



\bibitem{Metsaev:1999ui}
  R.~R.~Metsaev,
  Nucl.\ Phys.\ B {\bf 563}, 295 (1999)
  [arXiv:hep-th/9906217].




\bibitem{Metsaev:2008fs}
R.~R.~Metsaev, Phys.\ Rev.\ D {\bf 78} 106010 (2008); arXiv:0805.3472
[hep-th].


\bibitem{Fronsdal:1978vb}
  C.~Fronsdal,
  Phys.\ Rev.\ D {\bf 20}, 848 (1979).



\bibitem{Lopatin:1987hz}
  V.~E.~Lopatin and M.~A.~Vasiliev,
  Mod.\ Phys.\ Lett.\ A {\bf 3}, 257 (1988).

\bibitem{Vasiliev:1987tk}
  M.~A.~Vasiliev,
  Nucl.\ Phys.\ B {\bf 301}, 26 (1988).



\bibitem{Labastida:1987kw}
  J.~M.~F.~Labastida,
  Nucl.\ Phys.\ B {\bf 322}, 185 (1989).


\bibitem{Bekaert:2006ix}
  X.~Bekaert and N.~Boulanger,
  Commun.\ Math.\ Phys.\  {\bf 271}, 723 (2007)
  [arXiv:hep-th/0606198].


\bibitem{Hallowell:2005np}
  K.~Hallowell and A.~Waldron,
  Nucl.\ Phys.\ B {\bf 724}, 453 (2005)
  [arXiv:hep-th/0505255].



\bibitem{Zinoviev:2001dt}
Yu.~M.~Zinoviev, ``On massive high spin particles in (A)dS,''
arXiv:hep-th/0108192.


\bibitem{Metsaev:2006zy}
  R.~R.~Metsaev,
  Phys.\ Lett.\  B {\bf 643}, 205 (2006)
  [arXiv:hep-th/0609029].


\bibitem{Francia:2002aa}
  D.~Francia and A.~Sagnotti,
  Phys.\ Lett.\ B {\bf 543}, 303 (2002)
  [arXiv:hep-th/0207002].


\bibitem{Sagnotti:2003qa}
  A.~Sagnotti and M.~Tsulaia,
  Nucl.\ Phys.\  B {\bf 682}, 83 (2004)
  [arXiv:hep-th/0311257].




\bibitem{Buchbinder:2005ua}
  I.~L.~Buchbinder and V.~A.~Krykhtin,
  Nucl.\ Phys.\ B {\bf 727}, 537 (2005)
  [arXiv:hep-th/0505092].



\bibitem{Buchbinder:2006ge}
  I.~L.~Buchbinder, V.~A.~Krykhtin and P.~M.~Lavrov,
  Nucl.\ Phys.\  B {\bf 762}, 344 (2007)
  hep-th/0608005



\bibitem{Engquist:2007kz}
  J.~Engquist and O.~Hohm,
  Nucl.\ Phys.\  B {\bf 786}, 1 (2007)
  [arXiv:0705.3714 [hep-th]].


\bibitem{Buchbinder:2007ak}
  I.L. Buchbinder, A.V. Galajinsky and V.A.Krykhtin,
  Nucl.Phys.B {\bf 779}, 155 (2007)
  hep-th/0702161


\bibitem{Alkalaev:2003qv}
K.~B.~Alkalaev, O.~V.~Shaynkman and M.~A.~Vasiliev,
Nucl.\ Phys.\  B {\bf 692}, 363 (2004) [arXiv:hep-th/0311164].
%
%
arXiv:hep-th/0601225.



\bibitem{Skvortsov:2008vs}
  E.~D.~Skvortsov,
  JHEP {\bf 0807}, 004 (2008)
  [arXiv:0801.2268 [hep-th]];
``Frame-like Actions for Massless Mixed-Symmetry Fields in Minkowski space,''
  arXiv:0807.0903 [hep-th].


\bibitem{Iazeolla:2008ix}
  C.~Iazeolla and P.~Sundell,
  JHEP {\bf 0810}, 022 (2008)
  [arXiv:0806.1942 [hep-th]].



\bibitem{Brodsky:2008pg}
  S.~J.~Brodsky and G.~F.~de Teramond,
  arXiv:0802.0514 [hep-ph].



\bibitem{Andreev:2002aw}
O.~Andreev,
Phys.\ Rev.\  D {\bf 67}, 046001 (2003) [arXiv:hep-th/0209256].


\bibitem{Grigoryan:2007vg}
  H.~R.~Grigoryan and A.~V.~Radyushkin,
  Phys.\ Lett.\  B {\bf 650}, 421 (2007)
  [arXiv:hep-ph/0703069].






\bibitem{Metsaev:2007fq}
R.~R.~Metsaev, ``Ordinary-derivative formulation of conformal low spin
fields,'' arXiv:0707.4437 [hep-th].
%
%
``Ordinary-derivative formulation of conformal totally symmetric arbitrary
spin bosonic fields,'' arXiv:0709.4392 [hep-th].



\bibitem{Bolotin:1999fa}
  K.~I.~Bolotin and M.~A.~Vasiliev,
  Phys.\ Lett.\  B {\bf 479}, 421 (2000)
  [arXiv:hep-th/0001031].



\bibitem{Guttenberg:2008qe}
S.~Guttenberg and G.~Savvidy,
SIGMAP bulletin 4, 061 (2008) arXiv:0804.0522 [hep-th].


\bibitem{Manvelyan:2008ks}
  R.~Manvelyan, K.~Mkrtchyan and W.~Ruhl,
  Nucl.\ Phys.\  B {\bf 803}, 405 (2008)
  [arXiv:0804.1211 [hep-th]].




\bibitem{Buchbinder:2001bs}
  I.~L.~Buchbinder, A.~Pashnev and M.~Tsulaia,
  Phys.\ Lett.\  B {\bf 523}, 338 (2001)
  [arXiv:hep-th/0109067].


\bibitem{Tseytlin:2002gz}
  A.~A.~Tseytlin,
  Theor.\ Math.\ Phys.\  {\bf 133}, 1376 (2002)
  [arXiv:hep-th/0201112].



\bibitem{Bonelli:2003zu}
  G.~Bonelli,
  JHEP {\bf 0311}, 028 (2003)
  [arXiv:hep-th/0309222].



\bibitem{Deser:1983mm}
  S.~Deser and R.~I.~Nepomechie,
  Annals Phys.\  {\bf 154}, 396 (1984).


\bibitem{Deser:2003gw}
S.~Deser and A.~Waldron,
Nucl.\ Phys.\ B {\bf 662}, 379 (2003) [arXiv:hep-th/0301068].
%
Phys.\ Rev.\ Lett.\  {\bf 87}, 031601 (2001) [arXiv:hep-th/0102166].


\bibitem{Deser:2001us}
  S.~Deser and A.~Waldron,
  Nucl.\ Phys.\ B {\bf 607}, 577 (2001)
  [arXiv:hep-th/0103198].
%
  Phys.\ Lett.\  B {\bf 603}, 30 (2004)
  [arXiv:hep-th/0408155].



\bibitem{Skvortsov:2006at}
  E.~D.~Skvortsov and M.~A.~Vasiliev,
  Nucl.\ Phys.\  B {\bf 756}, 117 (2006)
  [arXiv:hep-th/0601095].




\end{thebibliography}
\end{document}